\pdfoutput=1 

\documentclass[11pt,journal]{IEEEtran}

  \usepackage{cite}

\usepackage[pdftex]{graphicx}
   \graphicspath{./img/}
   \DeclareGraphicsExtensions{.pdf,.jpeg,.png}
\usepackage{algorithm} 
\usepackage{algorithmic}

\usepackage{array}

\usepackage{mdwmath}
\usepackage{mdwtab}

\usepackage{graphicx}
\usepackage{caption}
\usepackage{subcaption}
  \usepackage[caption=false,font=footnotesize]{subfig}

\usepackage[cmex10]{amsmath}
\usepackage{amssymb} 
\usepackage{amsfonts}
\usepackage{amsthm}

\usepackage{url}

\newcommand{\eg}{{\it e.g.}}  \newcommand{\ie}{{\it i.e.}}
 \newcommand{\reals}{{\mathbb R}}
 
\newcommand\plusvee{\mathrel{\ooalign{\lower.5ex
\hbox{$\scriptstyle\vee\mkern.5mu$}\cr\hidewidth\raise.450ex
\hbox{$\scriptstyle+$}\cr}}}

\newcommand{\minimize}{\operatornamewithlimits{minimize}}

\begin{document}
%
\title{LocDyn: Robust Distributed Localization for Mobile Underwater
  Networks}
%
%
%
%

\author{Cl\'{a}udia Soares,~\IEEEmembership{Member,~IEEE,}
        Jo\~{a}o Gomes,~\IEEEmembership{Member,~IEEE,}
        Beatriz Ferreira,~\IEEEmembership{Student Member,~IEEE,}
        and~Jo\~{a}o Paulo Costeira
\thanks{This research was partially supported by EU-H2020 WiMUST project (grant agreement No. 645141) and Funda\c{c}\~{a}o para a Ci\^{e}ncia e Tecnologia (project UID/EEA/50009/2013).}}

\maketitle

\begin{abstract}
  How to self-localize large teams of underwater nodes using only
  noisy range measurements? How to do it in a distributed way, and
  incorporating dynamics into the problem? How to reject outliers and
  produce trustworthy position estimates? The stringent acoustic
  communication channel and the accuracy needs of our geophysical
  survey application demand faster and more accurate localization
  methods. We approach dynamic localization as a MAP estimation
  problem where the prior encodes dynamics, and we devise a convex
  relaxation method that takes advantage of previous estimates at each
  measurement acquisition step; The algorithm converges at an optimal
  rate for first order methods. LocDyn is distributed: there is no
  fusion center responsible for processing acquired data and the same
  simple computations are performed for each node. LocDyn is accurate:
  experiments attest to a smaller positioning error than a comparable
  Kalman filter. LocDyn is robust: it rejects outlier noise, while the
  comparing methods succumb in terms of positioning error.
\end{abstract}

\begin{IEEEkeywords}
Range-based localization, Distributed localization, Autonomous underwater
vehicles, Mobile location estimation, Robust network localization.
\end{IEEEkeywords}


%

\section{Introduction}
\label{sec:introduction}

The development of networked systems of agents that can interact with
the physical world and carry out complex tasks in various contexts is
currently a major driver for research and technological development~\cite{NRC2015}.  This trend is also seen in contemporary ocean
applications and propelled research projects on multi-vehicle systems
like MORPH (Kalwa et al.~\cite{kalwa2013morph}) and, recently, WiMUST
(Al-Khatib et al.~\cite{al2015widely}).

Coordinated operation of vehicles requires a communication network to
share data, most critically, those data related to navigation and
positioning, as further explored in Abreu et
al.~\cite{AbreuBayatBotelhoGoisGomesPascoal2015}. Our work concerns
localization of (underwater) vehicles, a key subsystem needed in the
absence of GPS to properly georeference any acquired data and also
used in cooperative control algorithms. This paper presents research
results within the scope of EU H2020 project WiMUST, aiming at
advanced control, communication and signal processing tools to enable
a team of marine robots, either on the surface or submerged, to
jointly conduct geoacoustic surveys.

Today, geophysicists reveal sub-bottom structures using powerful sound
sources and hydrophones. During surveys, a towed source produces
acoustic waves that penetrate the sea bottom, and its layers
are inferred from the pattern of echoes observed at the towed
hydrophones, over a long period of time and a wide geographic
area. Such surveys are routinely carried out to characterize the sea
bottom prior to underwater construction, to monitor pipelines and
submerged structures, and for the operation of offshore oil and gas
fields.
\begin{figure}[t]
  \centering
  \includegraphics[width=\columnwidth]{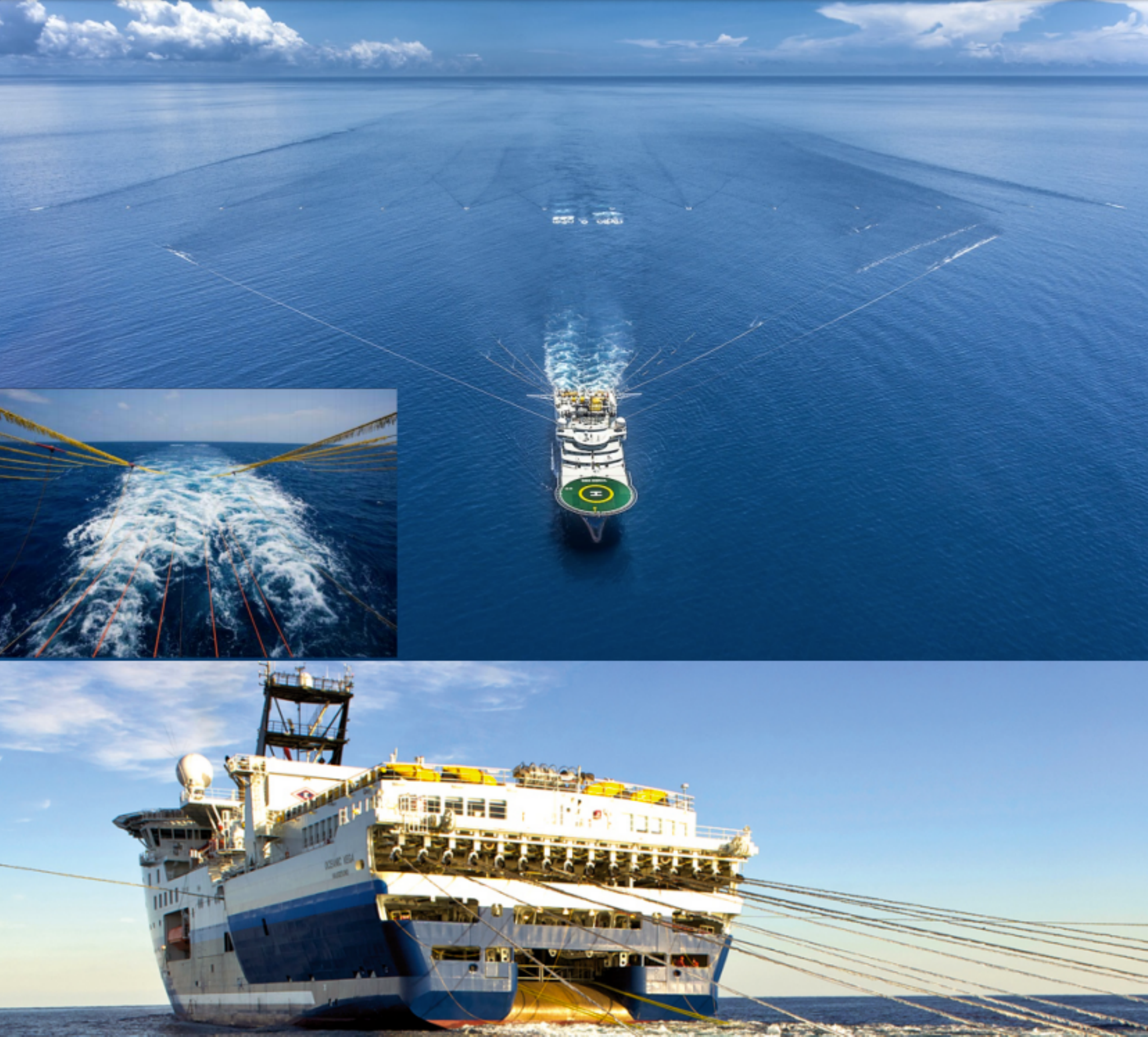}
  \caption{A surface vessel tows streamers with acoustic receivers for
    geophysical surveys (Al-Khatib et al.~\cite{al2015widely}).}
  \label{fig:big-towing-vessel}
\end{figure}
As depicted in Figure~\ref{fig:big-towing-vessel}, a single vessel
tows very long arrays of streamers and, thus, operation of a
traditional geophysical survey at sea means we cannot change
trajectories to recheck interesting findings; also, maneuvering
between rectilinear transects while keeping the streamers untangled
is challenging. The vision of WiMUST is to replace the monolithic
setup with a more flexible one where multiple heterogeneous underwater
vehicles tow smaller arrays while retaining a precise spatial
alignment. These are easier to maneuver, and the absence of long
physical ties between the surface ship and the data acquisition
devices enables new capabilities such as operating at variable depths
or adaptively changing the shape of the ensemble of
hydrophones.

Self-localization is a cornerstone for multi-vehicle cooperative
control in general and for WiMUST in particular, as the acoustic
signals must be georeferenced to high precision to enable an accurate
inference of deep sub-bottom layers. Our specific goal in this paper
is to accurately localize a network of moving agents from noisy
inter-vehicle ranges and from the positions of a few anchors or
landmarks.

\paragraph*{Related work}

The signal processing and control communities studied the network
localization problem in many variants, like static or dynamic network
localization, centralized or distributed computations,
maximum-likelihood methods, approximation algorithms, or outlier
robust methods.  

The control community's mainstream approach to localization
relies on the robust and strong properties of the Kalman
filter to dynamically compensate noise and bias. Recent approaches can
be found in Pinheiro et al.~\cite{PinheiroMorenoSousaRodriguez2016}
and Rad et al.~\cite{RadWaterschootToonLeus2011}. In the first very
recent paper, position and velocity are estimated from ranges,
accelerometer readings and gyroscope measurements with an Extended
Kalman filter. The authors of the second paper linearize the dynamic
network localization problem, solving it with a linear Kalman filter.
This last method is comparable with our range-only problem, although
the method requires knowledge of the noise's standard deviation.

The signal processing community traditionally studies static network
localization from a centralized perspective, like Keller and
Gur~\cite{KellerGur2011}, that formulate the problem as a regression
over adaptive bases; but the authors use squared distances, prone to
outlier noise amplification. Shang et
al.~\cite{ShangRumiZhangFromherz2004} follow a multidimensional
scaling approach, but multidimensional scaling works well only in
networks with high connectivity --- a property not encountered in
practice in large-scale geometric networks. Biswas et
al.~\cite{BiswasLiangTohYeWang2006} and more recently O\u{g}uz-Ekim et
al.~\cite{OguzGomesXavierOliveira2011} proposed semi-definite and
second order cone relaxations of the maximum likelihood
estimator. Although more precise, these convexified problems get
intractable even for a small number of nodes. Recently, we have
witnessed increasing interest from signal processing in distributed
static network localization. Papers by Shi et
al.~\cite{ShiHeChenJiang2010}, Srirangarajan et
al.~\cite{SrirangarajanTewfikLuo2008}, Chan and So~\cite{ChanSo2009},
Khan et al.~\cite{KhanKarMoura2010}, Simonetto and
Leus~\cite{SimonettoLeus2014} and recently Soares et
al.~\cite{soares2014simple} use different convex approximations to the
nonconvex optimization costs to devise scalable and distributed
algorithms for network localization. But for scenarios where
approximate solutions are not enough, researchers optimized the
maximum likelihood directly, obtaining solutions that depend on the
initialization of the algorithm. The methods in Calafiore et
al.~\cite{CalafioreCarloneWei2010} and Soares et
al.~\cite{SoaresXavierGomes2014a} increase the precision of a
relaxation-based solution, but are prone to local minima if wrongly
initialized. Lately, signal processing researchers produced solutions
for dynamic network localization; Schlupkothen et
al.~\cite{SchlupkothenDartmannAscheid2015} incorporated velocity
information from past position estimates to bias the solution of a
static localization problem via a regularization term.

\paragraph*{Our approach}

In this paper we deal with the network localization problem from an
optimization-based standpoint.  We formalize the network localization
problem under the maximum \textit{a posteriori} framework considering
white Gaussian noise and we tightly relax the nonconvex estimator to a
convex unconstrained program. We optimize the approximated problem
with a scalable and fast first order method, achieving smooth
trajectories for a small number of distributed iterations. We define
distributed operation as requiring no central or fusion node, and
where all nodes perform the same types of computations. Distributed
operation of vehicles requires the existence of a communication
network to share navigation and positioning data.

We propose a distributed algorithm for network localization of
underwater mobile nodes with the main properties of following a
principled maximum a posteriori approach, distributed iterations at
each agent, robustness to outlier measurements, and fast convergence.
We call our algorithm localization under dynamics, or LocDyn for
short.

While LocDyn supports distributed operation in the classic sense, it
may also be viewed in a more restricted way simply as an efficient
parallel algorithm when run at a central location that collects all
required range measurements through an appropriate forwarding protocol
(discussed, \eg, in Ludovico et
al.~\cite{LudovicoGomesAlvesFurfaro2014}). Depending on the capacity of
the shared transmission medium the latter solution may be preferable
from a practical standpoint, but it does not impact the derivations
below.

Our approach is most closely related to one-shot network localization
methods in signal processing (i.e., starting anew when repeated over
time), but it adds a temporal dimension that enables filtering to
regularize position estimates and thus improve their
accuracy. Contrary to many Kalman-filtering-based approaches to
localization in the control literature, we assume an extremely simple
dynamic model for mobile nodes and do not rely on navigation
information that could be provided by an inertial measurement unit
(IMU). The rationale for this is that mobile nodes in our scenarios
are not necessarily AUVs; provisions could be made to install some
range-measurement and communication devices in streamers, whose
dynamics are not as well characterized as those of the towing AUVs,
and where the presence of high-quality IMUs seems far-fetched at
present.

\subsection{Contributions}
\label{sec:contributions}
We introduce LocDyn, the first optimization-based dynamic network
localization estimator that is fully distributed and has optimal
convergence rate. LocDyn tightly approximates the MAP estimator, with
the nodes' dynamics as priors, so Bayesian estimation properties
are to be expected. We use a position predictor with information from
previous position estimates via a low-pass velocity approximation.
Our method is more accurate than a Kalman filter implementation by
more than 30cm per trajectory point in all our experiments.

In our companion UCOMMS'16 paper (Ferreira et al.~\cite{FerreiraGomesSoaresCosteira2016}) we focused on demonstrating
benefits for collaborative localization using hybrid range/bearing
measurements and time-domain filtering. We explore the same
fundamental idea for time-recursive processing here, but the method
for predicting velocities is now considerably improved, and we propose
an efficient distributed localization algorithm, whereas in
Ferreira et al.\ the dynamic optimization
problem was solved using a general-purpose (centralized) convex
solver. Also, the algorithm is carefully characterized and benchmarked
using numeric simulations.

The hybrid setup of Ferreira et al.
adopts the FLORIS/CLORIS least-squares framework (from a previous work
by Ferreira et al.~\cite{FerreiraGomesCosteira2015}), which in turn relies
on a so-called disk-based relaxation presented in Soares at
al.~\cite{soares2014simple} to attain a high-precision convex
formulation that is amenable to distributed/parallel processing. In
the present paper we consider exclusively range measurements to
streamline the technical content, but we emphasize that accommodating
bearing measurements in a hybrid localization scheme involves only
minor adaptations in the optimization problem and distributed solution
algorithm.

\section{Static network localization}
\label{sec:nlp}
The network of range-measurement and communication devices (nodes),
installed on AUVs and conceivably on streamers as well, is represented
as an undirected connected graph
$\mathcal{G} = (\mathcal{V},\mathcal{E})$.  The node set
$\mathcal{V} = \{1,2, \dots, n\}$ denotes the agents with unknown
positions. There is an edge $i \sim j \in {\mathcal E}$ between $i$
and $j$ if a noisy range measurement between nodes $i$ and $j$ is
available at both, and if $i$ and $j$ can communicate with each other.
The set of landmarks with known positions\footnote{These are
  considered constant for convenience, but could also be surface-bound
  mobile devices with permanently known positions obtained through
  GPS.}, named anchors, is denoted by
${\mathcal A} = \{ 1, \ldots, m \}$. For each $i \in {\mathcal V}$, we
let ${\mathcal A}_i \subset {\mathcal A}$ be the subset of anchors (if
any) relative to which node $i$ also possesses a noisy range
measurement.

Let $\reals^p$ be the space of interest ($p=2$ for planar networks,
and $p=3$ in the volumetric scenarios of greater interest here),
$x_i \in \reals^p$ the position of sensor $i$, and $d_{ij}$ the noisy
range measurement between sensors $i$ and $j$, known by both $i$ and
$j$. Without loss of generality, we assume $d_{ij} = d_{ji}$. Anchor
positions are denoted by $a_{k} \in \reals^{p}$. Similarly, $r_{ik}$
is the noisy range measurement between sensor $i$ and anchor $k$,
available at sensor $i$. 

The distributed network localization problem addressed in this work
consists in estimating the sensors' positions $x = \{ x_i\, : \, i \in
\mathcal{V} \}$, from the available measurements $\{ d_{ij} \, : \, i
\sim j \} \cup \{ r_{ik} \, : \, i \in {\mathcal V}, k \in {\mathcal
A}_i \}$, through collaborative message passing between neighboring
agents in the communication graph~${\mathcal G}$.

Under the assumption of zero-mean, independent and
identically-distributed, additive Gaussian measurement noise, the
maximum-likelihood estimator for the nodes' positions is the solution
of the optimization problem
\begin{equation}
  \label{eq:snlOptimizationProblem} 
  \minimize_{x} f(x),
\end{equation} 
where
\begin{equation*} 
  f(x) = \sum _{i \sim j} \frac{1}{2}(\|x_{i} - x_{j}\| - d_{ij})^2 + \sum_{i}
  \sum_{k \in \mathcal{A}_{i}} \frac{1}{2}(\|x_{i}-a_{k}\| - r_{ik})^2.
\end{equation*} 
Problem~\eqref{eq:snlOptimizationProblem} is nonconvex and difficult
to solve.  Even in the centralized setting (\ie, all measurements are
available at a central node) currently available iterative techniques
don't claim convergence to the global optimum. Also, even with
noiseless measurements, multiple solutions might exist due to
ambiguities in the network topology
itself~\cite{AspnesErenGoldenbergAnderson2006}.

We can address this problem by optimizing a convex approximation
to~\eqref{eq:snlOptimizationProblem}, amenable to distributed
implementation, as in Soares et al.~\cite{soares2014simple}. The
convex approximation~$\hat{f}$ is tight at each term of~$f$ and can be
optimized by a first order method with optimal convergence rate. The
approximated problem is
\begin{equation}
  \label{eq:cvx-static-problem}
  \minimize_{x} \hat{f}(x).
\end{equation}
The convex surrogate function~$\hat f$ is defined as
\begin{equation}
  \label{eq:hat-f}
  \hat{f}(x) = \sum_{i \sim j} \frac 12
  \mathrm{d}^{2}_{B_{ij}}(x_{i}-x_{j}) + \sum_{i} \sum_{k \in
    \mathcal{A}_{i}} \frac 12 \mathrm{d}^{2}_{Ba_{ik}}(x_{i}),
\end{equation}
where~$\mathrm{d}^{2}_{B_{ij}}$ and~$\mathrm{d}^{2}_{Ba_{ik}}$ are the
squared distances to a ball~$B_{ij} = \{y : \|y\| \leq d_{ij}\}$, and
a ball~$Ba_{ik} = \{y : \|y-a_{k}\| \leq r_{ik}\}$, respectively. The
convexification strategy underlying \eqref{eq:hat-f} is to relax
spheres in the constraint sets of squared distance functions to balls
(disks) $B_{ij}$, $Ba_{ik}$, hence the name \emph{disk-based
  relaxation}. In the next section we will use function~$\hat{f}$ and
a modified version of problem~\eqref{eq:cvx-static-problem} to
localize underwater moving nodes.

\subsection{Assumptions}
\label{sec:assumptions}

Range-only position estimation needs at least
$p+1$ anchors, or an equivalent set of physical constraints, to avoid
spatial ambiguities~\cite{AspnesErenGoldenbergAnderson2006}.

Consequently, all range-only methods assume that the number
of anchors, or landmarks, is greater than the dimension of the
deployment space --- 3 anchors for planar deployment and 4 for a
volumetric one.

\section{Motion-aware localization}
\label{sec:motion-loc}

One naive approach to localize a network of moving agents would be to
estimate the vehicles' positions solving~\eqref{eq:cvx-static-problem}
at each time step. Although it is possible, it does not take advantage
of the knowledge of previously estimated positions, so something
will be lost in processing time or communication bandwidth.

To bring motion into play we invoke the concept of \emph{prior
  knowledge} in Bayesian statistics and assume a Gaussian prior on the
nodes' positions, now understood as random variables. Each position's
distribution depends on the Gaussian distribution of the noisy range
measurements, its own prior and distributions of neighboring nodes'
positions.
The prior is the predicted position
\begin{equation}
  \label{eq:predicted-position}
  \tilde{x}_{i}(k+1) = \hat{x}_{i}(k) + v(k)_{i} \Delta T,
\end{equation}
where~$\hat{x}_{i}(k)$ is the estimated position at time
step~$k$, and~$v_{i}(k)$ is the measured or estimated velocity of~$i$.
As measurements are not taken continuously, we model time in discrete
steps~$t = k \Delta T$, where~$t$ is continuous time,~$k$ is the time
step, and~$\Delta T$ is the sampling period. Without loss of
generality we consider~$\Delta T$ fixed.
The distance measurement between vehicles~$i$ and~$j$ at positions
$x_{i}^{\star}$ and $x_{j}^{\star}$ is modeled as
\begin{equation}\label{eq:dij}
  d_{ij} = \|x_{i}^{\star} - x_{j}^{\star} \| + \mathcal{N}(0,\sigma^{2}),
\end{equation}
and, similarly, the range measurement between vehicle~$i$ and
anchor~$k$ is
\begin{equation}
  r_{ik} = \|x_{i}^{\star} - a_{k} \| + \mathcal{N}(0,\sigma^{2}).\label{eq:rik}
\end{equation}
For each node $i$ the prior distribution is also Gaussian, centered
on~$\tilde{x}_{i}$ with variance $\varsigma^{2}$. Assuming
independency, the posterior distribution of the positions at a given
time step is, up to a normalization constant, $p(x|\{d\}) \propto
p(\{d\}|x)p(x)$. This evaluates to
\begin{equation*}
  \begin{array}[t]{rcl}
  p(x|\{d\}) &\propto& \prod _{i \sim j} p(\|x_{i}-x_{j}\| - d_{ij})\\
  &&\prod_{i} \left(\prod_{k \in \mathcal{A}} p(\|x_{i}-a_{k}\| -
    r_{ik}) p(x_{i} - \tilde{x}_{i})\right),
  \end{array}
\end{equation*}
where all densities on the right-hand side are Gaussian.
We cast network localization as a maximum \textit{a posteriori}
estimation problem. After applying the logarithm, we get
\begin{equation}
  \label{eq:map-prob}
  \minimize_{x} \frac 1{\sigma^{2}} f(x) + \frac 1{\varsigma^{2}}\sum_{i}\|x_{i}
  - \tilde{x}_{i}\|^{2},
\end{equation}
equivalently written as
\begin{equation*}
  \minimize_{x} f(x) + \lambda \|x - (\hat{x} + v \Delta T)\|^{2},
\end{equation*}
where we multiplied by~$\sigma^{2}$ and
adopted~$\lambda = \frac {\sigma^{2}} {\varsigma^{2}}$. Thus, the
parameter~$\lambda$ has a physical interpretation: it is the ratio of
the uncertainty in the measurements to the richness of the trajectory.
The concatenated vehicles'
velocities at time~$k$, $v(k)$, can be measured or approximated from
the previous location estimates.
As we have seen, this problem is nonconvex, so we convexify it using
the approach from section~\ref{sec:nlp} to obtain the problem
\begin{equation}
  \label{eq:penalty-problem}
  \minimize_{x} g_{\lambda}(x) = \hat{f}(x) + \lambda \|x - (\hat{x} +
  v \Delta T)\|^{2}.
\end{equation}
We can also interpret~\eqref{eq:penalty-problem} as a regularized
network localization problem. The regularization parameter~$\lambda$
controls how much we want to bias our estimate towards the predicted
position.
\begin{figure}[t]
  \centering
  \includegraphics[width=0.7\columnwidth]{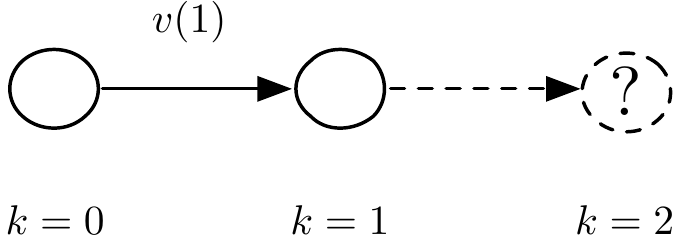}
  \caption{Localization of a moving vehicle: at time~$k=2$ our
    formulation uses velocity information from the previous estimates
    and biases the estimator in~\eqref{eq:penalty-problem} towards the
    dotted position.}
  \label{fig:localization-example}
\end{figure}
For example, if a node moves linearly as in
Figure~\ref{fig:localization-example}, at~$k=2$ our formulation will
use velocity~$v(1)$ to predict the position of the vehicle, and bias
the static localization problem towards the predicted solution.

We introduced problem~\eqref{eq:penalty-problem} in
Ferreira et al.~\cite{FerreiraGomesSoaresCosteira2016} in the context
of hybrid collaborative localization based on range and bearing
measurements.

\section{The LocDyn algorithm}
\label{sec:locdyn}

Problem~\eqref{eq:penalty-problem} has properties that allow fast
optimization: it is a sum of convex functions and, thus, convex. It is
actually strongly convex\footnote{Recall that a function~$g$ is
  $m$-strongly convex if and only if $g(x) - \frac m2 x^{\top}x$ is
  convex for all~$x$.}, meaning that at any point~$x$ the function is
lower bounded by a quadratic and thus possesses a unique minimum on
compact sets (c.f.\ Boyd and
Vandenberghe~\cite{BoydVandenberghe2004}).  Also, our objective
function in~\eqref{eq:penalty-problem} is L-smooth, meaning that there
is a quadratic upper bound to~$g_{\lambda}(x)$ for all~$x$,
so~$g_{\lambda}$ does not grow too
fast. Appendix~\ref{sec:l-smoothn-g_lambda} holds proofs and
computation of the $L$ and $m$ constants.
If~$g_{\lambda}$ is strongly convex and has a Lipschitz continuous
gradient, a first-order minimization algorithm can be maximally
accelerated (c.f.\ Nesterov~\cite{Nesterov2004}).

The gradient of~$g_{\lambda}$ is
\begin{equation}
  \label{eq:grad}
  \nabla g_{\lambda} = \nabla \hat f (x) + 2 \lambda (x-\tilde x),
\end{equation}
where~$\nabla \hat{f}$ is, as defined in~(15) of Soares et
al.~\cite{soares2014simple}:
\begin{equation}
  \label{eq:grad-f-hat}
  \nabla \hat f(x) = \mathcal{L}x -A^{\top}\mathrm{P_{B}}(Ax) +
  \begin{bmatrix}
    \sum_{k \in \mathcal{A}_{1}}x_{1} - \mathrm{P_{Ba}}_{1k}(x_{1})\\
    \vdots\\
    \sum_{k \in \mathcal{A}_{n}}x_{n} - \mathrm{P_{Ba}}_{nk}(x_{n})
   \end{bmatrix},
\end{equation}
where~$A=C \otimes I_{p}$, $C$ is the arc-node incidence matrix of
$\mathcal{G}$,~$I_{p}$ is the identity matrix of size~$p$, and
$\mathrm{B}$ is the Cartesian product of the
balls~$\mathrm{B}_{ij} = \{y : \|y\| \leq d_{ij}\}$ corresponding to
all the edges in~$\mathcal{E}$. Similarly,
$\mathrm{Ba}_{ik} = \{y : \|y-a_{k}\| \leq r_{ik}\}$. Also,
$\mathcal{L}= A^{\top}A = L \otimes I_{p}$, with $L$ being the
Laplacian matrix of graph~$\mathcal{G}$.

\begin{algorithm}[tb]
  \caption{LocDyn}
  \label{alg:LocDyn}
  \begin{algorithmic}[1] 
    \REQUIRE $
    \begin{array}[t]{l}
      L; \\
      \{d_{ij}(k) : i \sim j \in \mathcal{E}\};\\
      \{r_{ik}(k) : i \in \mathcal{V}, k \in \mathcal{A}\};\\
      \beta = \frac{1-\sqrt{m/L}}{1+\sqrt{m/L}}; \\
    \end{array}
$
    \ENSURE $\hat x(k)$
    \STATE measure or estimate (via~\eqref{eq:fir-velocity})
    $v_{i}(k-1) \Delta T$
    \STATE $\tilde{x}_{i} = \hat{x}_{i}(k-1) + v_{i}(k-1) \Delta T$;
    \STATE $\kappa = 0;$
    \STATE each node $i$ chooses random $x_{i}(0) = x_{i}(-1)$;
    \WHILE{some stopping criterion is not met, each node $i$}
    \STATE $\kappa = \kappa+1$
    \STATE $
    \begin{aligned}[t]
      w_{i} =
      x_{i}(\kappa-1)+\beta\left(x_{i}(\kappa-1)-x_{i}(\kappa-2)\right);
    \end{aligned}
    $ \label{alg:extrapolated-point}
    \STATE node $i$ broadcasts $w_{i}$ to its neighbors 
    \STATE $
    \begin{array}{l}
      \nabla g_{\lambda i}(w_{i}) = \delta_{i}w_{i} - \sum_{j \in N_{i}}w_{j}
      + \\
      + \sum_{j \in N_{i}} c_{(i \sim j,i)} \mathrm{P_{B}}_{ij}(w_{i}
      - w_{j}) +  2 \lambda (w_{i}-\tilde{x}_{i});
    \end{array}
    $ \label{alg:gradg}
    \STATE $
    \begin{aligned}[t]
      x_{i}(\kappa) = w_{i} - \frac{1}{L}\nabla g_{\lambda i}(w_{i});
    \end{aligned}
    $ \label{alg:updatex}
    \ENDWHILE
    \RETURN $\hat{x}(k) = x(\kappa)$
  \end{algorithmic}
\end{algorithm}  
Algorithm~\ref{alg:LocDyn} specifies LocDyn as detailed this far, with
its regularization term. As discussed previously,~$k$ indexes time
steps where we have range data and anchor positions acquisition. in
the interval, we run the algorithm, whose steps are indexed
by~$\kappa$. The procedure inherits and concurs with the
distributed properties of the static method in Soares et al.\
\cite{soares2014simple}. Step~\ref{alg:extrapolated-point} computes
the extrapolated points~$w_i$ in a standard application of Nesterov's
method~\cite{Vandenberghe2014FastProxGrad}.  Step~\ref{alg:gradg}
corresponds to the $i$-th entry of~$\nabla \hat f$ and an affine term
on~$x_{i}$ dependent only on each node's unknown coordinates,
velocity, and the position estimated in the previous time
step. Constants $c_{(i \sim j,i)}$ denote the entry $(i \sim j,i)$ in
the arc-node incidence matrix $C$, and~$\delta_{i}$ is the degree of
node~$i$. The $i$-th entry of~$\mathcal{L}x$ can be computed by
node~$i$ from its current position estimate and the position estimates
of the neighbors,
as~$(\mathcal{L}x)_{i} = \delta_{i}x_{i} - \sum_{j \in
  N_{i}}x_{j}$. As further detailed in Soares et
al.~\cite{soares2014simple},
\begin{equation*}
  (A^{\top}\mathrm{P_{B}}(Ax))_{i} = \sum_{j \in N_{i}}c_{(i \sim
  j,i)} \mathrm{P_{B}}_{ij}(x_{i}-x_{j}),
\end{equation*}
as presented in Step~\ref{alg:gradg}.

Next, we deal with how to approximate the velocity of a node in the
global reference frame only with data collected so far.

\subsection{Velocity estimation}
\label{sec:velocity-estimation}
To include vehicle dynamics in network localization, we penalize
discrepancies between the predicted and estimated positions. The
predicted positions are computed based on the previous estimated
position and the vehicle's velocity in world coordinates. This
velocity can be measured by the AUV, or, as we consider next, it can
be estimated from the moving pattern so far. In a previous paper
(Ferreira et al.~\cite{FerreiraGomesSoaresCosteira2016}), inspired by
the recent work by Schlupkothen et
al.~\cite{SchlupkothenDartmannAscheid2015}, we estimated the velocity
of each vehicle in the global reference frame by averaging the norm
and the angle over a sliding time window. But prediction by averaging
is accurate only if the averaged quantities are nearly constant
through time --- meaning linear constant motion. This is not
necessarily the case in the projected futuristic scenarios for
geoacoustic surveying; in this type of application, richer
trajectories such as the one depicted in Figure~\ref{fig:lawnmower1}
are meant to densely cover the geographic area under study. In this
paper we estimate each vehicle's velocity~$\hat{v}_{i}$ by taking
Taylor expansions of the derivative of the position.  We start by
approximating velocity by central finite differences. Unlike the
causal backward Euler difference approximation, that converges
linearly, the centered difference approximation converges
quadratically as~$\Delta T \to 0$ and is defined as:
\begin{equation*}
 \hat{v}_{i}(k) \Delta T = \frac{x_{i}(k+1) - x_{i}(k-1)}{2}.
\end{equation*}
Higher order approximations have even faster convergence rates, and
they are more robust to noise in the position estimates. Nevertheless,
to use them in causal estimators like LocDyn, we have to introduce a
time lag that covers the higher order time shifts. As communication in
the underwater acoustic channel has low bandwidth and slow propagation
speed, networking protocols may entail considerable latency that
invalidates the estimation of velocities with too large time shifts, so there
is a tradeoff between accuracy and opportunity in choosing the order
of approximation. A causal sixth-order approximation is
\begin{equation}
  \begin{array}[t]{rcl}
    \hat{v}_{i}(k) \Delta T &=& \frac{45 (x_{i}(k-3)-x_{i}(k-5))}{60} + \\
                            &&- \frac{9 (x_{i}(k-2) - x_{i}(k-6))}{60} + \\
                            &&\frac{x_{i}(k-1)-x_{i}(k-7)}{60}.
  \end{array}\label{eq:taylor-velocity}
\end{equation}
However, computing differences amplifies noise. To reduce the impact
of noise in velocity estimation, we take the derivative approximation
as an anti-symmetric FIR filter with a defined accuracy order. The
derivative approximation's transfer function can then be designed to
match the transfer function of the continuous derivative
operator. Holoborodko~\cite{Holoborodko2008} proposed a method to
generate differentiator filters that also cut high frequencies,
rejecting noise both in measurements and in the estimation process,
while preserving the differentiation behavior at low frequencies. This
method was successfully used, for example, by Khong et
al.~\cite{khong2013} or Hosseini and
Plataniotis~\cite{hosseini2014}. We use the smooth low noise
differentiator fitted from a second-degree polynomial with a lag of~7
samples. Although it takes the same estimates
as~\eqref{eq:taylor-velocity}, the low-pass designed coefficients
reject high-frequency noise. The expression for the product between
estimated velocity and sampling is
\begin{equation}
\begin{array}{lcl}
  \hat{v}_{i}(k) \Delta T &=& \frac {5 (x_{i}(k-3) - x_{i}(k-5))}{32} + \\
                          &&\frac{4 (x_{i}(k-2)-x_{i}(k-6))}{32} + \\
                          &&\frac{x_{i}(k-1)-x_{i}(k-7)}{32}.
\end{array}\label{eq:fir-velocity}
\end{equation}

\subsection{Convergence}
\label{sec:alg-convergence}
The accelerated gradient method implemented in Alg.~\ref{alg:LocDyn}
for function~$g_{\lambda}$ with constants~$m$ and~$L$ specified
in~\eqref{eq:m} and~\eqref{eq:L} converges at the optimal rate
$O\left( \kappa^{-2} \right)$ as proved by Nesterov
\cite{Nesterov1983},\cite{Nesterov2004}. Also, the distance to the
unique global optimum~$g_{\lambda}^\star$ at iteration~$\kappa$ is
theoretically bounded by
\begin{equation*}
  \begin{array}[t]{l}
g_{\lambda}( x(\kappa) ) - g_{\lambda}^\star \leq \\
\frac{4}{(2+\kappa\sqrt{m/L})^2} \left( g_{\lambda}(x(0)) -
  g_{\lambda}^\star + \frac m2 \left\| x(0) - x^\star
  \right\|^2\right).
  \end{array}
\end{equation*}

\section{Numerical evaluation}
\label{sec:experimental-evaluation}

\begin{figure}[t]
    \centering
    \begin{subfigure}[b]{0.3\textwidth}
        \includegraphics[width=\textwidth]{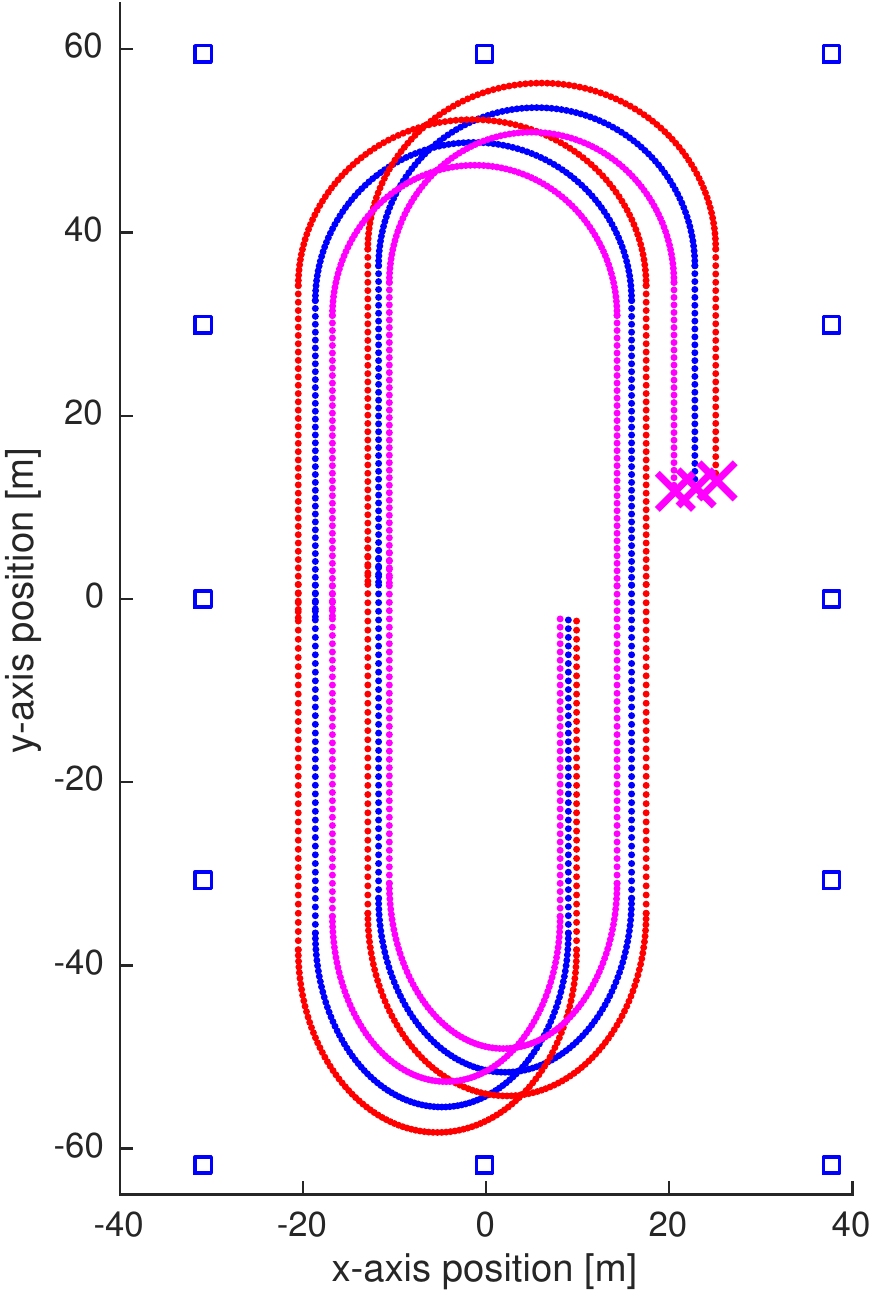}
        \caption{Lap trajectory, with three vehicles (distances in meters).}
        \label{fig:lapExample0}
    \end{subfigure}
    ~ 
    \begin{subfigure}[b]{0.3\textwidth}
        \includegraphics[width=\textwidth]{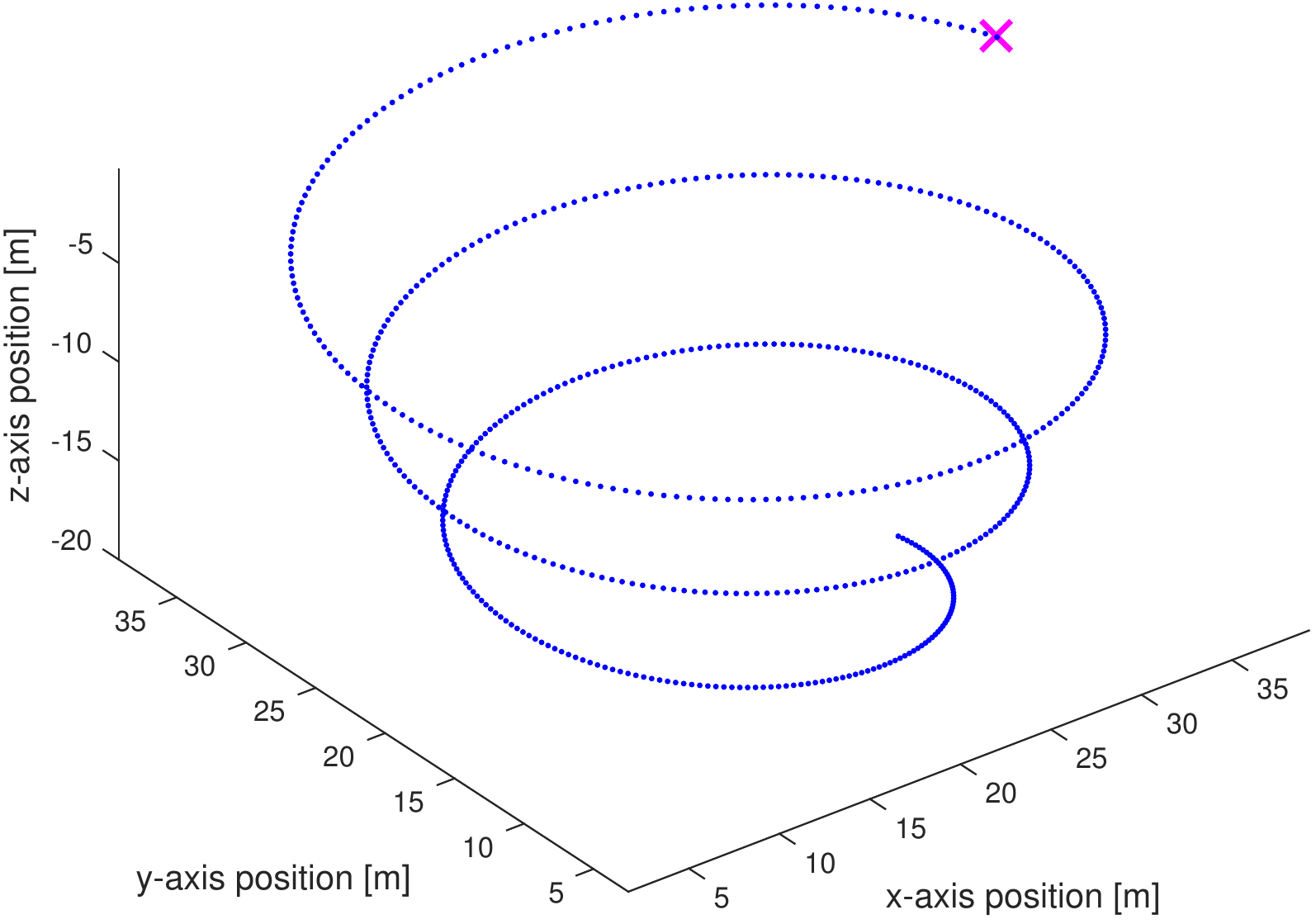}
        \caption{Descending spiral.}
        \label{fig:spiralExample0}
    \end{subfigure}
    ~
    \begin{subfigure}[b]{0.3\textwidth}
        \includegraphics[width=\textwidth]{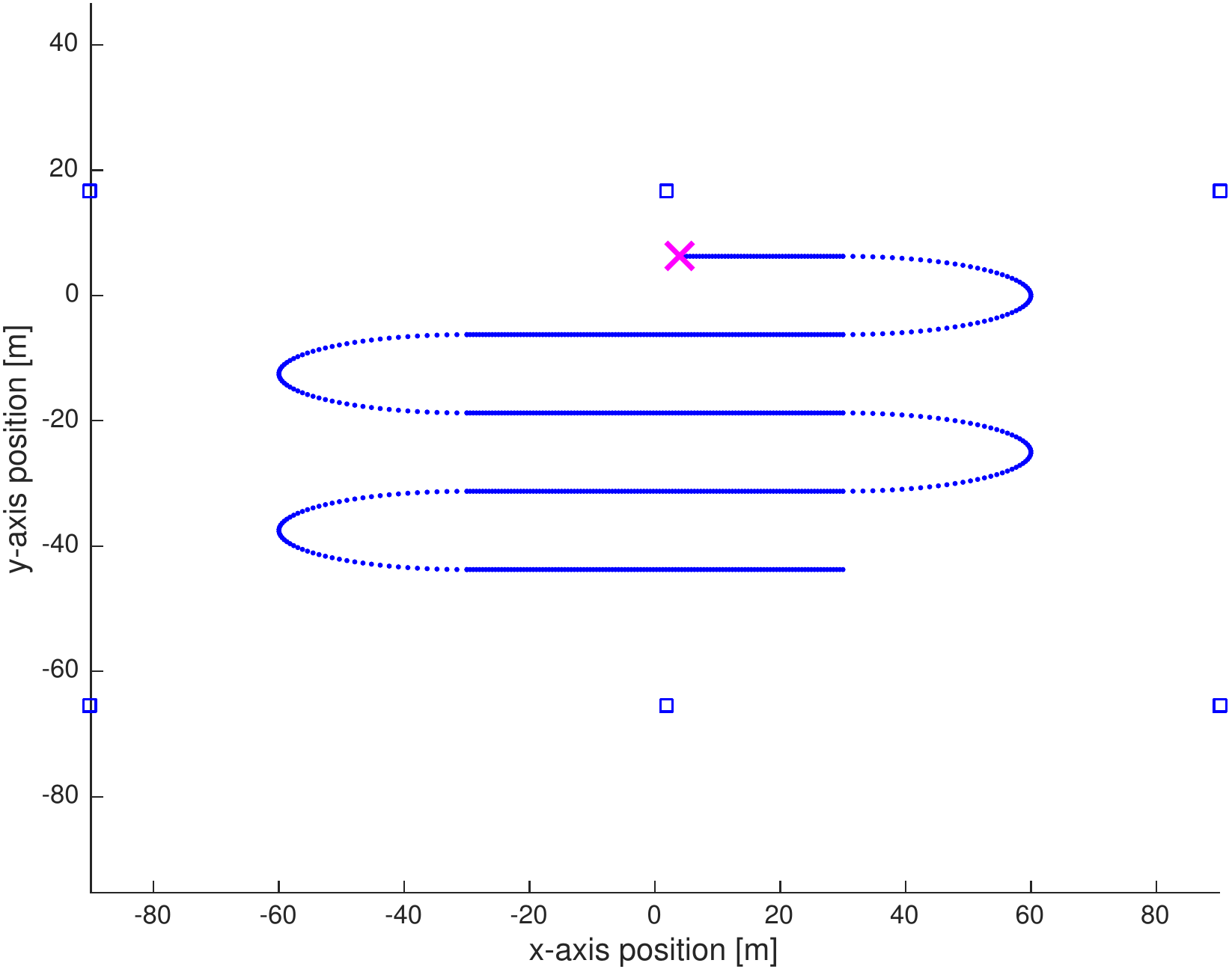}
        \caption{The lawn mower trajectory.}
        \label{fig:lawnmower1}
    \end{subfigure}
    \caption{Trajectories for experimental evaluation. The starting
      position is marked with a magenta cross. Anchors are depicted as
    blue squares.}\label{fig:trajectories}
\end{figure}
We evaluate LocDyn in trajectories used in (not necessarily
geoacoustic) surveys: a lap (Figure~\ref{fig:lapExample0}), descending
3D spiral (Figure~\ref{fig:spiralExample0}), and the lawn mower
(Figure~\ref{fig:lawnmower1}).

In all experiments, we contaminate distance measurements at each time
step~$k$ with zero-mean white Gaussian noise with standard deviation
~$\sigma = 1$m. This value for the standard deviation of
measurement noise is an upper bound on the real-world noise observed in the
WiMUST vehicles, where ranging errors are less than~1m.
We generated synthetic data according to
\begin{equation*}
  d_{ij} = |\|x_{i}^{\star} -x_{j}^{\star}\| + \mathcal{N}(0,\sigma^{2})|,
\end{equation*}
for inter-sensor distances and
\begin{equation*}
  r_{ik} = |\|x_{i}^{\star} -a_{k}\| + \mathcal{N}(0,\sigma^{2})|,
\end{equation*}
for sensor-anchor distances. We emphasize that measurements are in
mismatch with the data model considered in Sections~\ref{sec:nlp}
and~\ref{sec:motion-loc}, but the discrepancy is not serious, as the
likelihood of $d_{ij}$, $r_{ik}$ being nonpositive in~\eqref{eq:dij} and~\eqref{eq:rik} is typically very small.

We benchmark LocDyn against static localization in Soares et
al.~\cite{soares2014simple} and against the linear Kalman filter
solution proposed by Rad et al.~\cite{RadWaterschootToonLeus2011}. We
compare LocDyn with the Kalman filter in the lap and the lawn mower
trajectories and not in the spiral, because the
implementation that the authors kindly made available to us works only
in 2D. We provided the true~$\sigma$ to the Kalman filter, whereas
LocDyn and static localization do not use it. All the other Kalman
filter parameters were not altered.

We ran 100 Monte Carlo trials for each experiment, and measured the
empirical error as
\begin{equation*}
  \mathrm{Error} = \frac 1{K} \sum_{k=1}^{K}\|\hat{x}(k) -
  x^{\star}(k) \|,
\end{equation*}
where~$K$ is the total number of steps in the
trajectory, and~$x^{\star}(k)$ is the concatenation of the true
positions~$x_{i}^{\star}$.
We initialize LocDyn and the Kalman filter in the  position
marked with a magenta cross. Static localization doesn't require
initialization. Anchors are placed to contain the trajectory on their
convex hull. In 2d, we placed 6 anchors for the lawn mower and 12 for
the lap. For 3D, we used 16 anchors.

\subsection{Lap trajectory}
\label{sec:lap}

\begin{figure}[t]
  \centering
  \includegraphics[width=0.9\columnwidth]{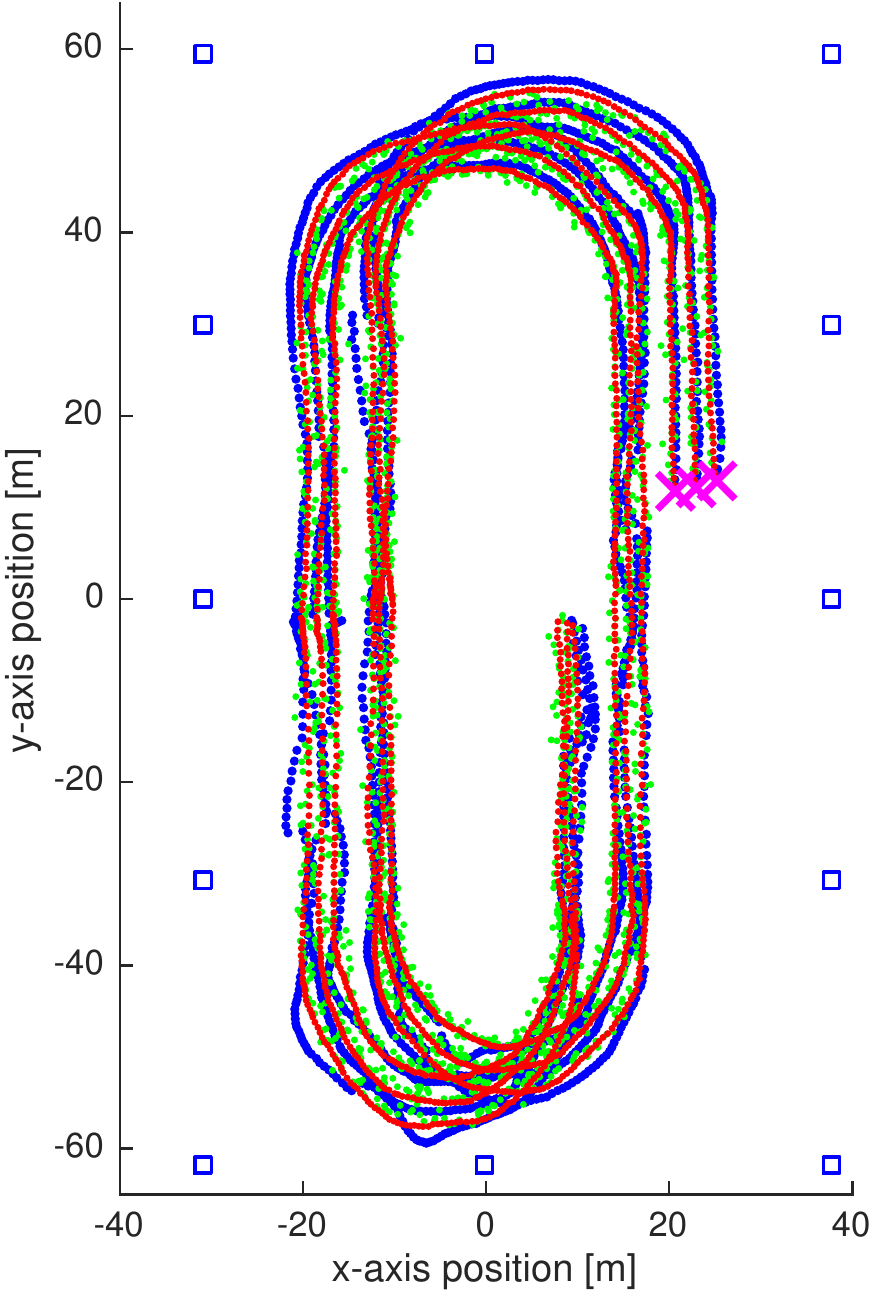}
  \caption{Example run: A team of three underwater vehicles
    describes a planar lap surveying trajectory. In green, we have the
    static localization estimates, in blue the Kalman filter, and in
    red, LocDyn. LocDyn describes the smoothest trajectory. There is a
    slowdown around (-15,-5) where the Kalman Filter gets lost, maybe
    due to the assumption of constant velocity.}
  \label{fig:lapExample}
\end{figure}
Lap trajectories are frequent in ocean surveying. They are also rich
since they combine linear parts with curved ones. We tested LocDyn,
the Kalman filter and static localization in the lap shown in
Figure~\ref{fig:lapExample0} and one of the Monte Carlo runs is
depicted in Figure~\ref{fig:lapExample}. We perceive the LocDyn
trajectory as the most natural of the three. There is an intentional
slowdown of the vehicles around (-15,-5) and LocDyn is the least
affected by it. We observed these behaviors in all our visualizations
of the lap estimated trajectories.
\begin{figure}[t]
  \centering
  \includegraphics[width=0.9\columnwidth]{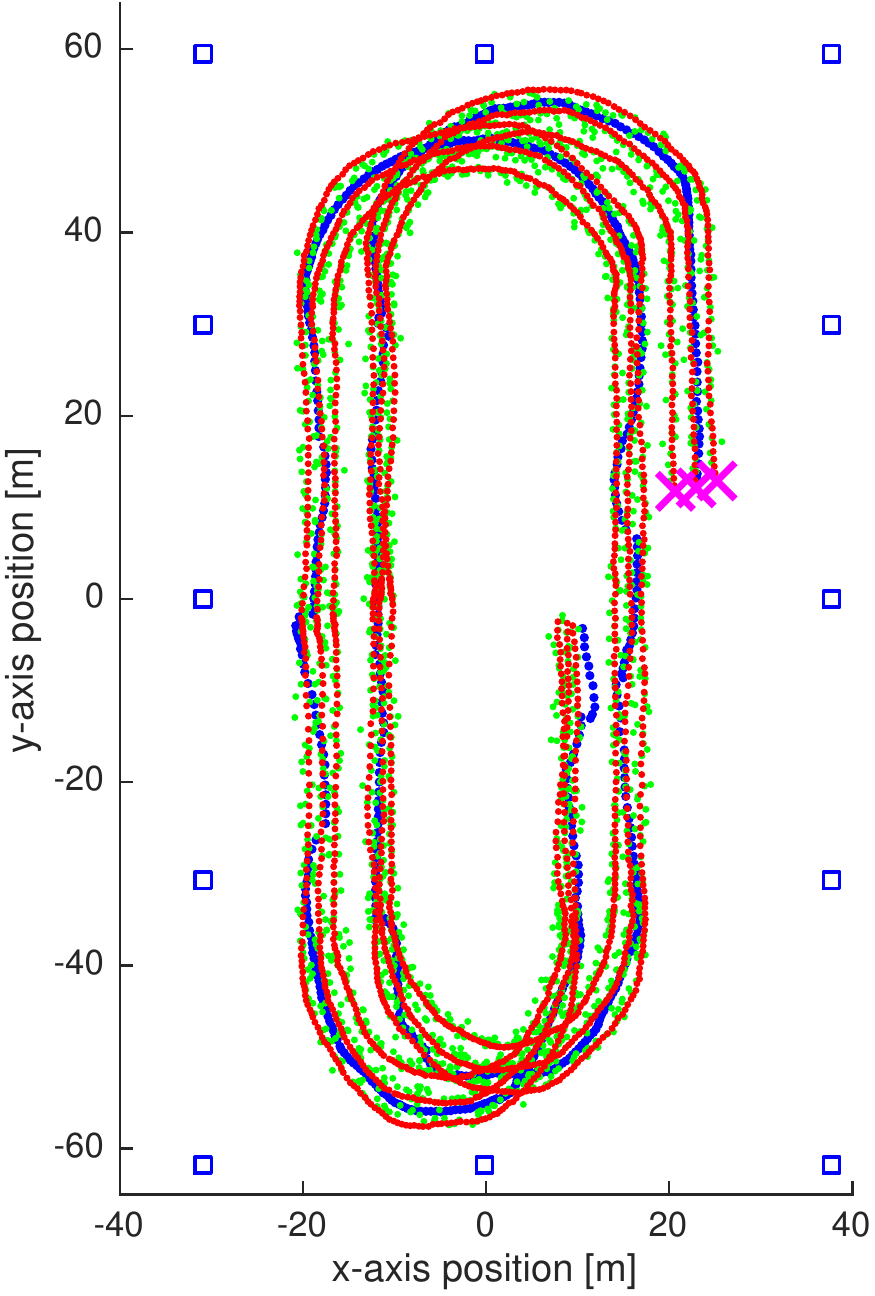}
  \caption{Example run: A team of three underwater vehicles
    describes a planar lap surveying trajectory. For the Kalman filter
  estimates, we display only the center vehicle's trajectory.}
  \label{fig:lapExample2}
\end{figure}
We can see in more detail the behavior of the Kalman filter estimates
in Figure~\ref{fig:lapExample2}, where we displayed only the center
vehicle. It transitions well from the linear to the circular part, but
it gets lost when entering the next linear one. When the first slowdown
starts it gets lost, and the same happens again, after the second
slowdown.
\begin{figure}[!t]
  \centering
  \includegraphics[width=0.9\columnwidth]{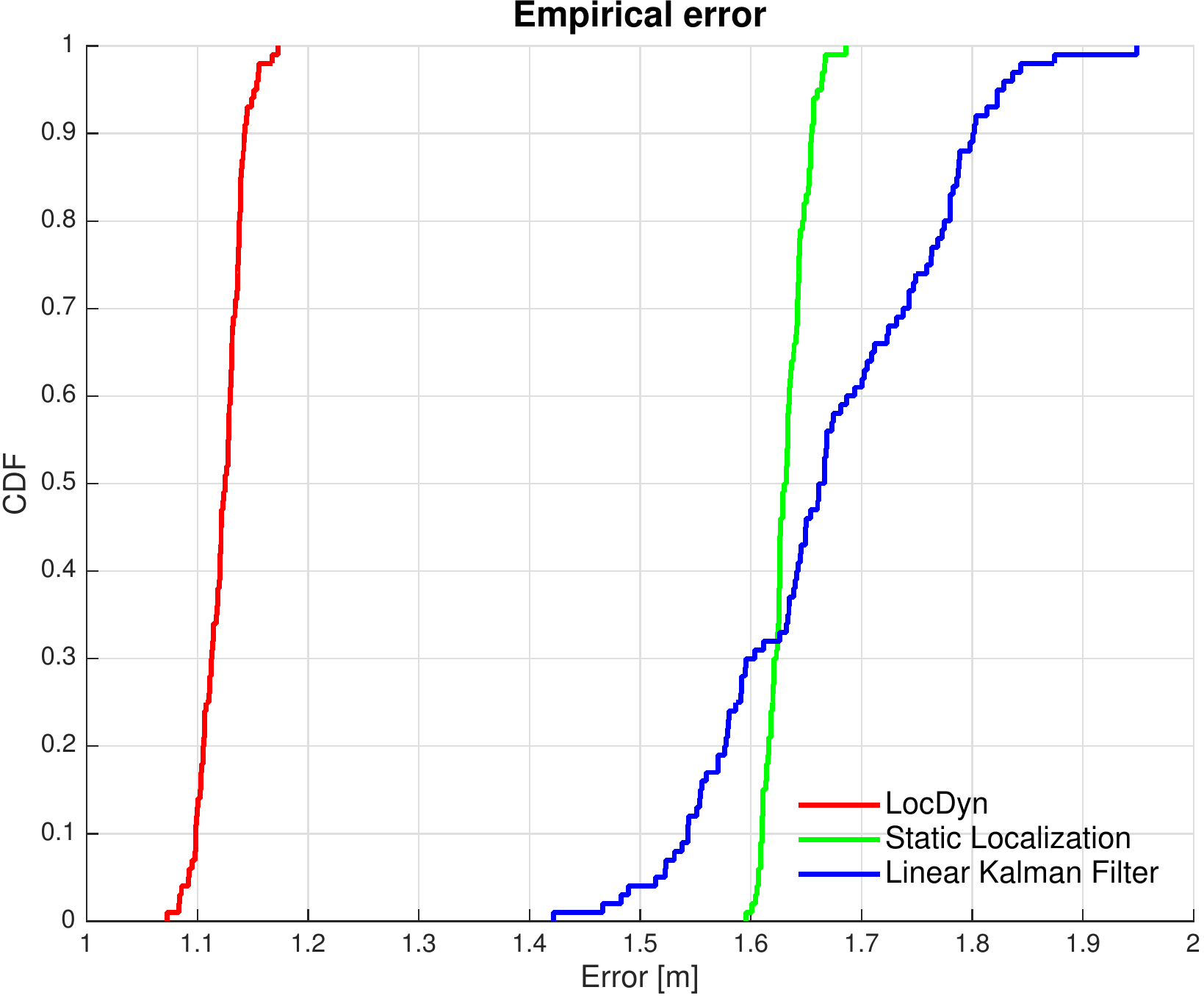}
  \caption{Empirical CDF for the planar lap trajectory with 100 Monte
    Carlo runs. LocDyn is the most accurate, and has a small variance
    in the empirical error. The Kalman filter fares the worst, mostly
    because of its slowdown behavior, as depicted in
    Figure~\ref{fig:lapExample2}.}
  \label{fig:fulllapMCCDF}
\end{figure}
But what about statistically relevant behavior? To answer this, we
simulated the lap for 100 Monte Carlo trials and computed the
empirical error of the trajectories. Figure~\ref{fig:fulllapMCCDF}
displays the resulting empirical cumulative distributions (CDF) for
the three algorithms. Not only is LocDyn more accurate, but it also
shows less variance in the error. The Kalman filter lags behind static
localization, although it delivers smoother trajectories. This
increase in the error is due to the bad accuracy near the slowdown
points discussed previously.

\subsection{Descending spiral trajectory}
\label{sec:desc-spir-traj}

Descending spirals are useful for monitoring the water column. They
are difficult trajectories to follow so, although they are not
associated with geophysical surveying activities, we tested LocDyn in
them.
\begin{figure}[t]
  \centering
  \includegraphics[width=0.9\columnwidth]{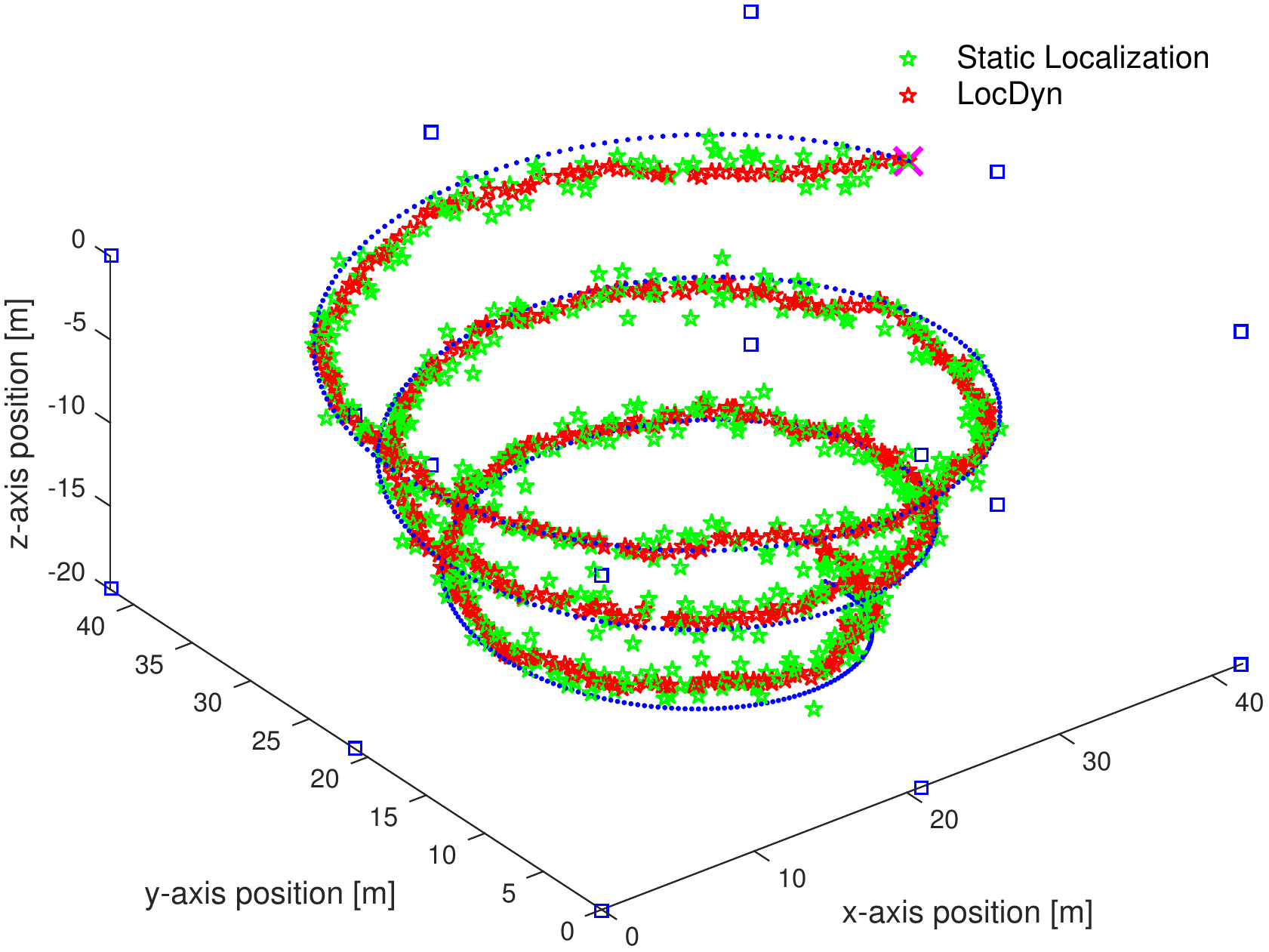}
  \caption{Example run: an AUV follows a descending spiral
    trajectory. Two algorithms attempt to localize it from noisy
    ranges and a few anchor positions. As expected, LocDyn has a
    natural trajectory and static localization scatters its estimates
    around the true trajectory.}
  \label{fig:fullspiralExample1}
\end{figure}
Figure~\ref{fig:fullspiralExample1} shows an example run of the
descending spiral trajectory. LocDyn has the smoothest trajectory,
although both LocDyn and static localization are more deviated towards
the center of the spiral. This is due to the small number of anchors
in this setup, considering the augmented degrees of freedom in passing
from 2D to 3D. Nevertheless, including dynamic information improves
drastically the localization accuracy.
\begin{figure}[t]
  \centering
  \includegraphics[width=0.9\columnwidth]{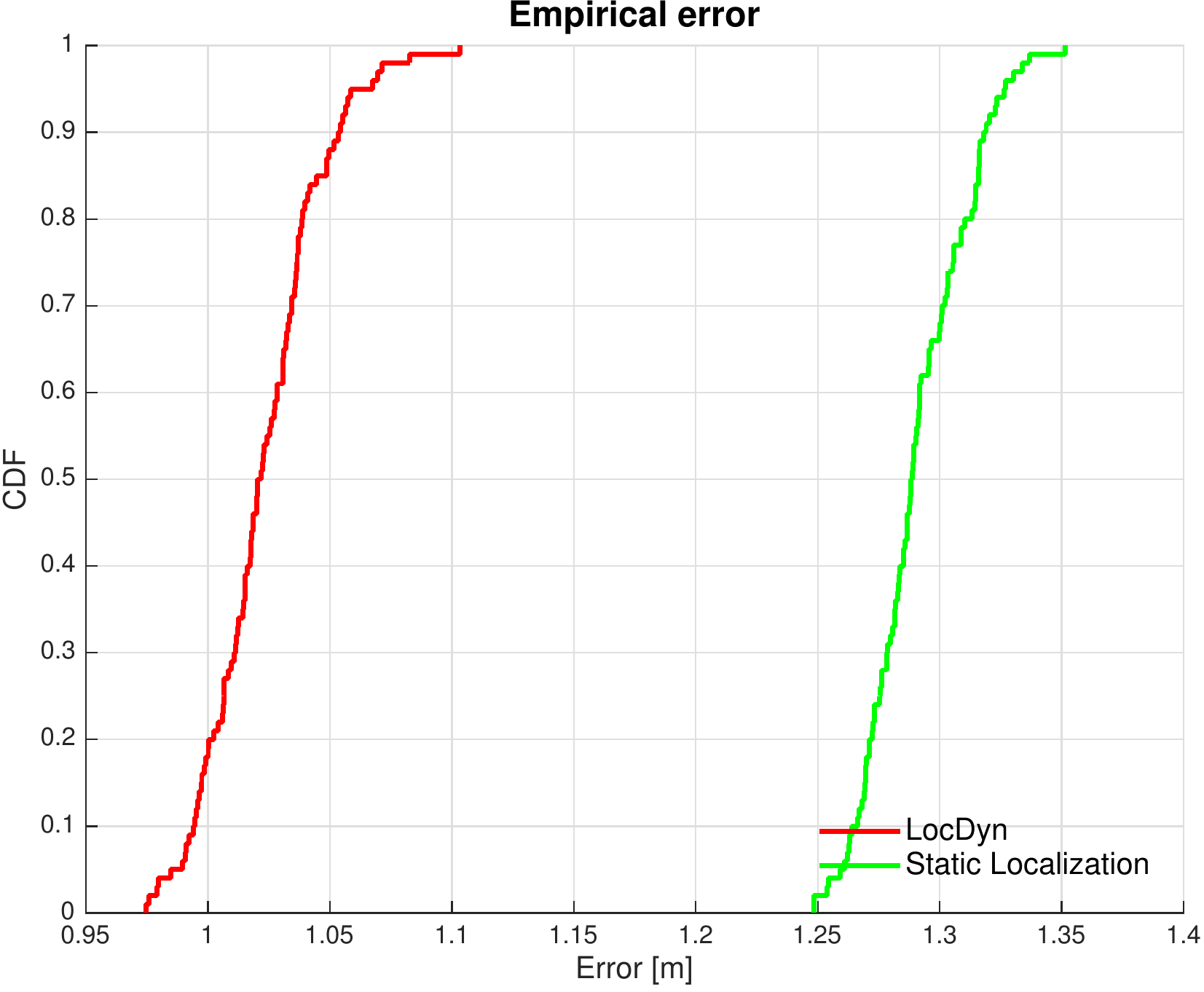}
  \caption{Empirical CDF for the descending spiral trajectory with 100
    Monte Carlo runs. LocDyn is the most accurate estimator.}
  \label{fig:fullspiralMCCDF}
\end{figure}
We documented the experimental accuracy increase in
Figure~\ref{fig:fullspiralMCCDF}. Here, we see the CDF of the
empirical error from 100 Monte Carlo simulations. Noticeably, LocDyn
confirms the intuition from the example trajectory of
Figure~\ref{fig:fullspiralExample1}: the average accuracy gain of
using LocDyn is of 30cm --- about one third of the 1m of measurement
noise standard deviation.

\subsection{Lawn mower trajectory}
\label{sec:lawn-mower-traj}

In this experiment we test the robustness of the algorithms to
outlier noise. Outliers are  due, for example, to
reflection or multi-path of the acoustic wave. At each step~$k$ with
probability~$1\%$ the noisy range measurement of the vehicle to the anchor
on the SW corner is doubled.
\begin{figure}[t]
  \centering
  \includegraphics[width=0.9\columnwidth]{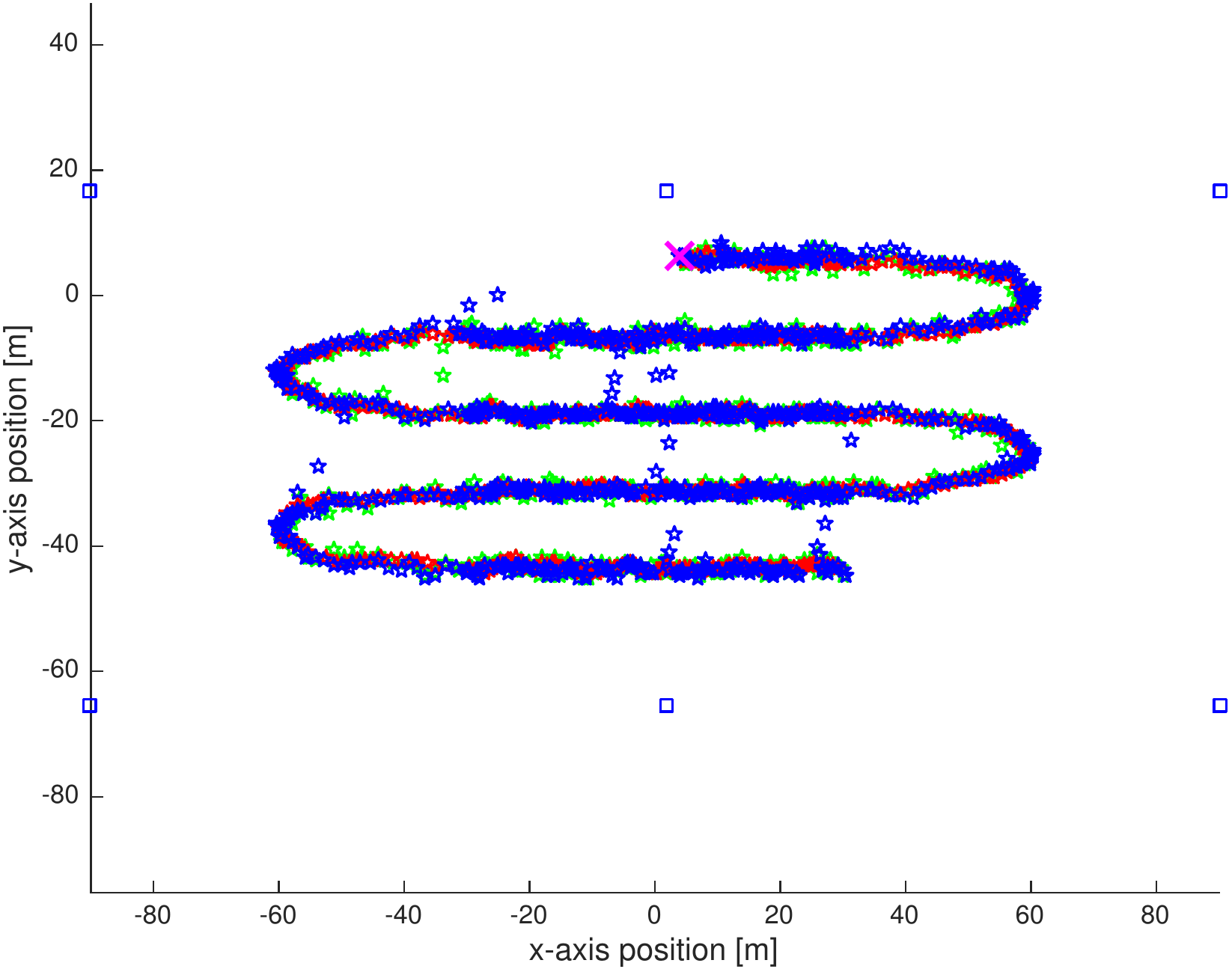}
  \caption{Example run: an AUV follows a lawn mower trajectory.}
  \label{fig:lawnmowerExample}
\end{figure}
Figure~\ref{fig:lawnmowerExample} depicts one example run of the
algorithms. At the outlier-contaminated steps we see that the Kalman
filter looses the lawn mower path. Less frequently, static
localization also increases the localization error.
\begin{figure}[t]
  \centering
  \includegraphics[width=0.9\columnwidth]{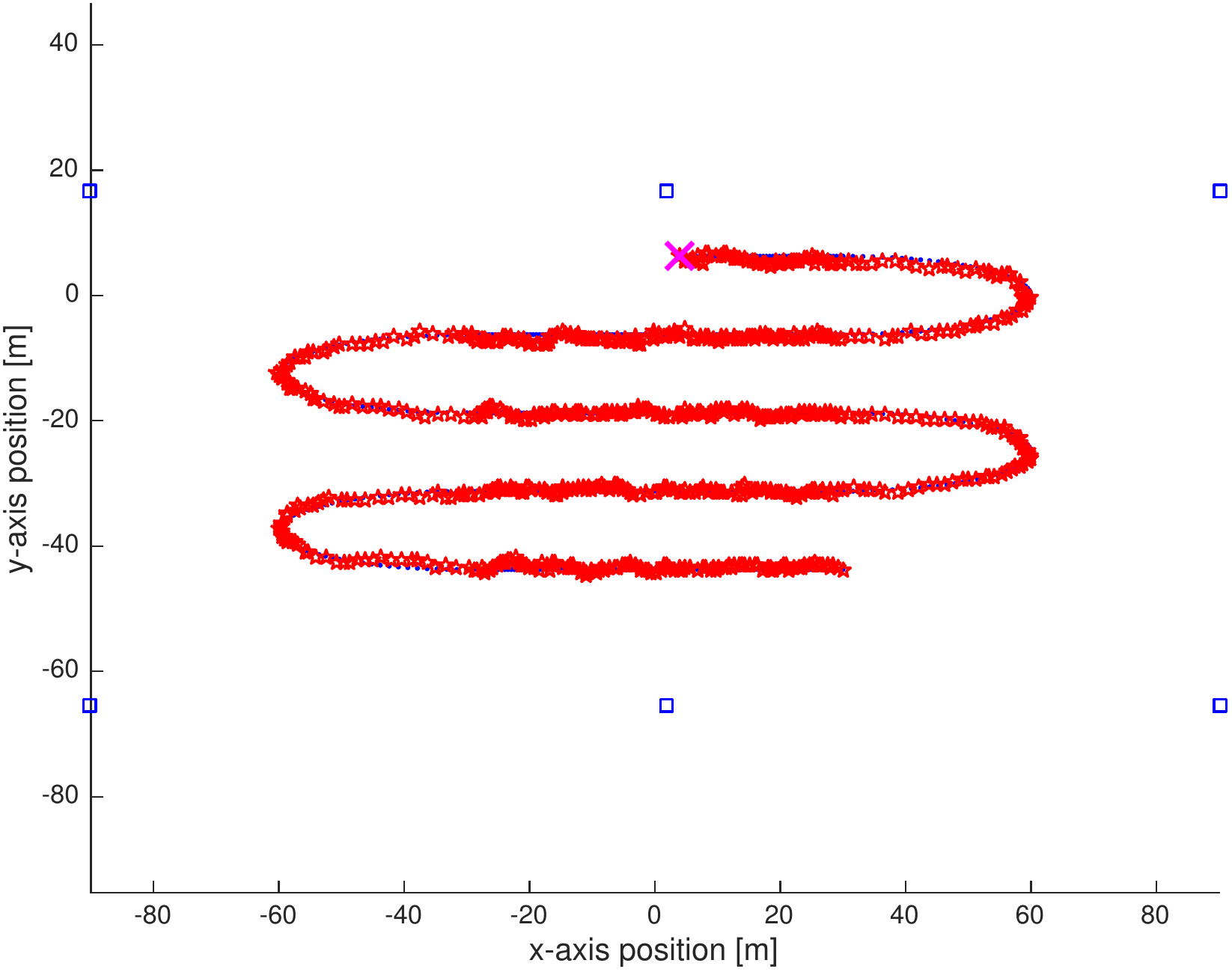}
  \caption{Example run: an AUV follows a lawn mower trajectory. LocDyn
  estimates.}
  \label{fig:lawnmowerLD}
\end{figure}
Figure~\ref{fig:lawnmowerLD} displays only LocDyn estimates, and
evidences that the trajectory was not particularly hurt.
\begin{figure}[t]
  \centering
  \includegraphics[width=0.9\columnwidth]{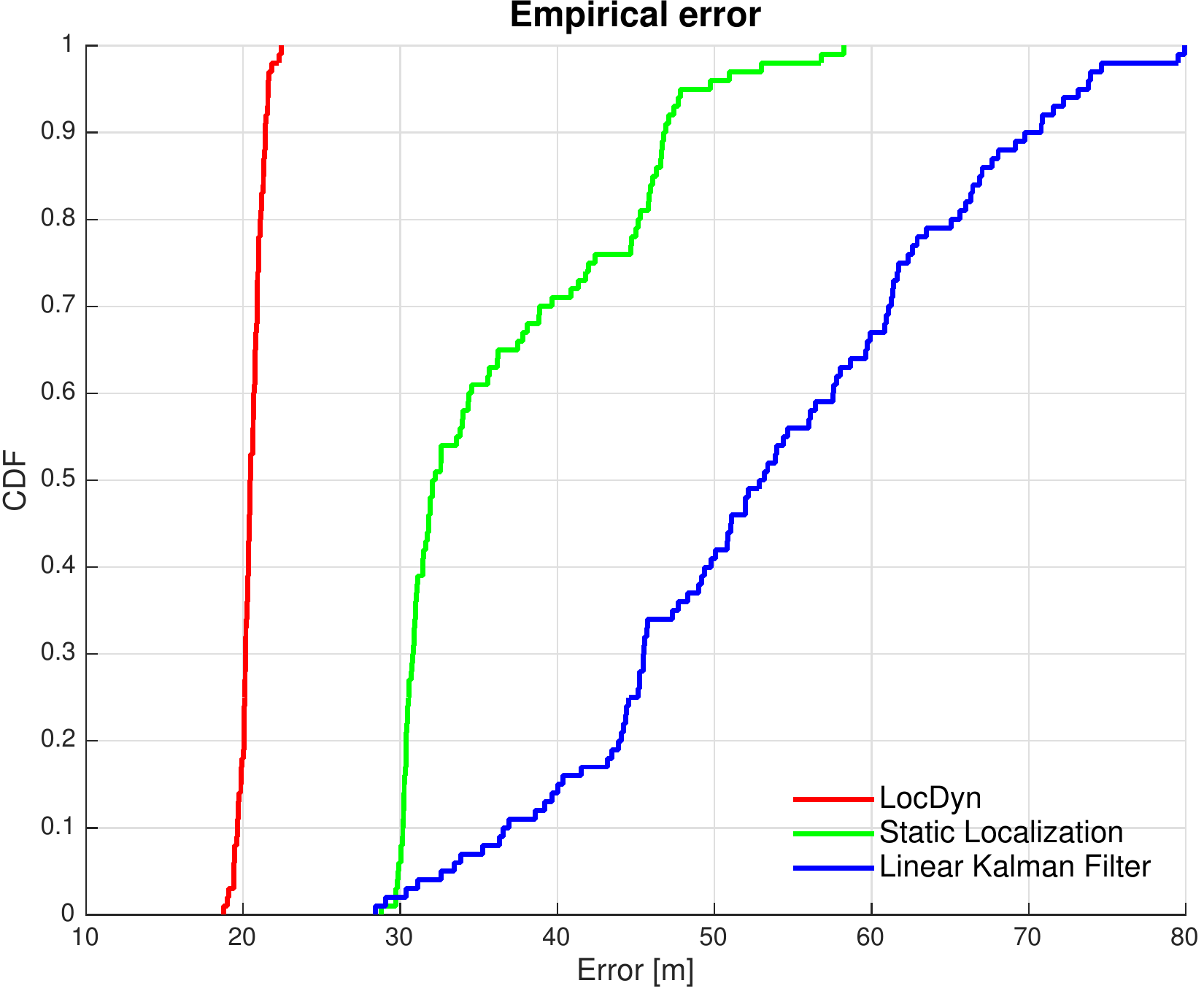}
  \caption{Empirical CDF for the lawn mower trajectory with 100
    Monte Carlo runs. LocDyn is the most accurate estimator.}
  \label{fig:fulllawnmowerMCCDF}
\end{figure}
The empirical CDF demonstrates the impact of outlier noise in the
positioning error, and confirms that LocDyn is not only the most
accurate by far, but also with much less variance of the error.

\section{Conclusion}
We explored self-localization of a network of underwater vehicles from
no more than noisy range measurements in a principled MAP estimation
framework. We produced a fast, and distributed algorithm, LocDyn,
topping in accuracy a comparable Kalman filter estimator. We showed
the advantage of encoding the dynamic behavior of moving vehicles, by
comparing LocDyn with a static network localization estimator.
Also, we gave physical meaning to LocDyn's only parameter, as a ratio
of variability in measurement noise to variability of trajectories. An
important open end of this work is to devise a way to eliminate the
parameter altogether so to have a parameter-free solution for
motion-aware network localization.


%



  \section*{Acknowledgment}

The authors would like to thank Ant\'{o}nio Pascoal and Jorge Ribeiro
from DSOR-ISR for the information regarding AUV missions, and Hadi
Jamali Rad, for kindly providing code for his method.

\appendices

\section{$L$-smoothness and $m$-strong convexity of~$g_{\lambda}$}
\label{sec:l-smoothn-g_lambda}

$L$-smoothness implies that the gradient of~$g_{\lambda}$ is Lipschitz
continuous with Lipschitz constant~$L$.  
A function~$g$ is Lipschitz continuous if
there exists a Lipschitz constant~$L$ such that
\begin{equation*}
  \|g(x) -g(y) \| \leq L \|x-y\|,
\end{equation*}
for all~$x$ and~$y$. We now prove that the gradient of~$g_{\lambda}$ is
Lipschitz continuous and a Lipschitz constant~$L$ can be identified by
\begin{IEEEeqnarray}{rCl}\IEEEnonumber
  \|\nabla g_{\lambda}(x) - \nabla g_{\lambda}(y) \|&=& \|\nabla
  \hat{f}(x) - \nabla \hat{f}(y) + 2 \lambda (x-y)\|\\ \IEEEnonumber
  &\leq& \|\nabla \hat{f}(x) - \nabla \hat{f}(y)\| + 2\lambda\|x-y\|\\ \label{eq:L}
  &\leq& \underbrace{(L_{\hat f} + 2\lambda)}_{L} \|x-y\|.
\end{IEEEeqnarray}
The inequality~\eqref{eq:L} refers to the constant~$L_{\hat f}$
in~(16) of Soares et al.~\cite{soares2014simple}.
A strong convexity
modulus~$m$ for~$g_{\lambda}$can be computed from
\begin{equation*}
  g_{\lambda}(x) - \frac m2 x^{\top}x  =  \hat{f}(x) + \left (
    \lambda - \frac m2 \right ) x^{\top}x -2 \lambda \tilde{x}^{\top}x
  + \lambda \tilde{x}^{\top}\tilde{x}
\end{equation*}
noticing that the left-hand side is convex if all terms in the
right-hand side are convex in~$x$. This entails that~$g_{\lambda}$ is
strongly convex with modulus
\begin{equation*}
  m \leq 2\lambda.
\end{equation*}
  As we want
robustness of~$g_{\lambda}$ to errors in~$x$, we choose~$m$ to have
the smallest condition number~$L/m$ for
function~$g_{\lambda}$; thus, from now on we take the largest
possible
\begin{equation}
  \label{eq:m}
  m = 2\lambda.
\end{equation}




%
\bibliographystyle{IEEEtran}
\bibliography{IEEEabrv,biblos.bib}

\end{document}